\documentclass[preprint,12pt]{aastex}

\usepackage{natbib}
\usepackage{graphicx}
\usepackage{subfigure}
\usepackage{color}
\usepackage[colorlinks,citecolor=blue]{hyperref}
\usepackage{mathrsfs}
\usepackage{rotating}

\newcommand{\src}{IGR~J17544$-$2619}
\newcommand{\xshalo}{X--ray scattered halo}
\newcommand{\xsring}{X--ray scattered ring}
\newcommand{\chd}{{\it Chandra}}
\newcommand{\xmm}{{\it XMM}--{\it Newton}}
\newcommand{\swi}{{\it Swift}}
\newcommand{\ein}{{\it Einstein}}
\newcommand{\ros}{{\it ROSAT}}
\newcommand{\suz}{{\it Suzaku}}
\newcommand{\intg}{{\it INTEGRAL}}

\slugcomment{Draft version \today}

\shorttitle{X--ray scattered halo around \src }
\shortauthors{Mao et al.}

\begin{document}

\title{X--ray Scattered Halo Around \src\ }

\author{Junjie Mao\altaffilmark{1,2},
Zhixing Ling\altaffilmark{1}, Shuang-Nan Zhang\altaffilmark{1,3,4}}

\email{zhangsn@ihep.ac.cn}

\altaffiltext{1}{National Astronomical Observatories, Chinese Academy of Sciences,
	Beijing 100012, China}
\altaffiltext{2}{University of Chinese Academy of Sciences,
	Beijing 100049, China}
\altaffiltext{3}{Key Laboratory of Particle Astrophysics, Institute of High Energy Physics,
	Beijing 100049, China}
\altaffiltext{4}{Physics Department, University of Alabama in
Huntsville, Huntsville, AL 35899, USA}

\begin{abstract}
X--ray photons coming from an X--ray point source not only
arrive at the detector directly,
but also can be strongly forward-scattered by the interstellar dust
along the line of sight (LOS),
leading to a detectable diffuse halo around the X--ray point source.
The geometry of small angle X--ray scattering is straightforward,
namely, the scattered photons travel longer paths and thus arrive later
than the unscattered ones;
thus the delay time of \xshalo\ photons can reveal information of
the distances of the interstellar dust and the point source.
Here we present a study of the \xshalo\ around \src ,
which is one of the so--called supergiant fast X--ray transients.
\src\ underwent a striking outburst when observed with \chd\ on 2004 July 3,
providing a near $\delta$--function lightcurve.
We find that the X--ray scattered halo around \src\ is
produced by two interstellar dust clouds along the LOS.
The one which is closer to the observer
gives the \xshalo\ at larger observational angles;
whereas the farther one, which is in the vicinity of the point source,
explains the halo with a smaller angular size.
By comparing
the observational angle of the scattered halo photons
with that predicted by different dust grain models,
we are able to determine the normalized dust distance.
With the delay times of the scattered halo photons,
we can determine the point source distance, given a dust grain model. Alternatively we can discriminate between the dust grain models,
if the point source distance is known independently.
\end{abstract}


\section{Introduction}
Immersed in the interstellar medium (ISM), the interstellar dust grains
not only absorb X--ray photons but also scatter X--ray photons.
\citet{ove65} first predicted the presence of \xshalo s around X--ray sources.
Nonetheless due to the limitation of the angular resolution
of X--ray imaging telescopes,
it was not until two decades later,
\citet{rol83} first observed the diffuse \xshalo\ around GX 339$-$4
with the Imaging Proportional Counter (IPC) instrument onboard \ein .
The theory of small angle X--ray scattering
has since been refined by a number of authors
\citep[e.g.,][etc]{mau86,mat91,smi98}.
Meanwhile, X--ray scattering phenomena around
both X--ray point sources and extended supernova remnants (SNRs)
were found with \ros\ \citep[e.g.,][]{pre95}, \chd\ \citep[e.g.,][]{smi02},
\xmm\ and \swi /XRT \citep[e.g.,][]{tie10}, etc.

Likewise, albeit it had already been pointed out by \citet{tru73} that
the time delay effect in the \xshalo s can be used to
determine the distances of variable X--ray point sources,
it was nearly three decades later then came its long overdue application.
\citet{pre00} roughly estimated the distance of Cyg X$-$3,
which was observed with the Advanced CCD Imaging Spectroscopy
(ACIS) instrument onboard \chd .
Thanks to the fine angular resolution of \chd ,
\citet{tho09},~\citet{lin09}, and~\citet{xia11}
used the time delay effect in the \xshalo s to
determine the distances of Cen X$-$3, Cyg X$-$3, and Cyg X$-$1, respectively.
Additionally, \citet{vau04} first expanded the scope onto $\gamma$--ray bursts (GRBs),
in which case an even simpler geometry was involved.
The evolving \xsring s around GRB~031203 \citep{vau04, tie06}, GRB~050724 \citep{vau06},
GRB~050713A \citep{tie06}, GRB~061019 and GRB~070129 \citep{via07}
were detected with both \xmm\ and \swi /XRT.

It is indisputable that the characteristics of
the X--ray dust scattered halos (e.g. the radial profile, etc)
depend upon the properties of interstellar dust grains,
including chemical composition, dust size distribution, dust spatial distribution,
and sometimes the morphology and
the alignment of elongated dust grains might play important roles.
So far, various types of interstellar dust grain models
have been established based on different observational results
such as interstellar extinction, diffuse infrared emission feature and so forth
\citep[see][for a review]{lia03}.
Among them, the three types of models provided by \citet[][hereafter MRN]{mat77},
\citet[][hereafter WD01]{wei01}, \citet[][hereafter ZDA]{zub04} are most widely used.

In this work we analyze the archival \chd\ data of \src ,
which was first discovered with \intg\ on 2003~September~17 \citep{sun03}
as a galactic high mass X--ray binary (HMXB).
With several subsequent observations carried out with
\xmm\ \citep{gri04}, \chd\ \citep{iza05},
\swi /XRT \citep{sid09}, and \suz\ \citep{ram09},
remarkable outbursts with a duration time scale of hours were spotted.
Along with the confirmation of an O9Ib blue supergiant donor
in the binary system \citep{pel06}, \src\ was confirmed as one of the so--called
supergiant fast X--ray transients \citep[SFXTs][]{sgu05,smi06,neg06}.
SFXT, a subclass of HMXB, is characterised by the presence of a supergiant companion
and significant outbursts lasting typically a few hours.
Typically, the peak luminosity of a flare
can be about a factor of 10$^{3}$--10$^{5}$ times the fainter quiescent X--ray luminosity.

In Section~\ref{sct:data}, we analyze the \chd\ ACIS-S data of \src ,
focusing on timing (Section~\ref{sct:timing}) and imaging analysis
(Section~\ref{sct:imaging}).
Then, we derive the time lags of the \xshalo\ photons
via cross correlation method in Section~\ref{sct:ccf}.
In Section~\ref{sct:theta},
we model the deviation of
the arithmetic mean of the observation angle
from the mid-value of the angular distance of the annular region.
We subsequently present the distance measurement for interstellar dust clouds
along the line of sight (LOS) and the point source
(Section ~\ref{sct:dist}).
A dynamical distance measurement for \src\,
obtained from a Galactic Center molecular clouds survey,
presented in Section \ref{sct:clouds_dist},
is consistent with the estimated point source distance.
In Section~\ref{sct:caveats}, we briefly discuss
the feasibility of a promising application of the relationship between
interstellar dust grain models and the average observational angles for annular halos.
Finally, we summarize our results in Section \ref{sct:summary}.

\section{Data Reduction and Analysis}
\label{sct:data}
\begin{figure}
\centering
\includegraphics[width=0.7\columnwidth,angle=0]{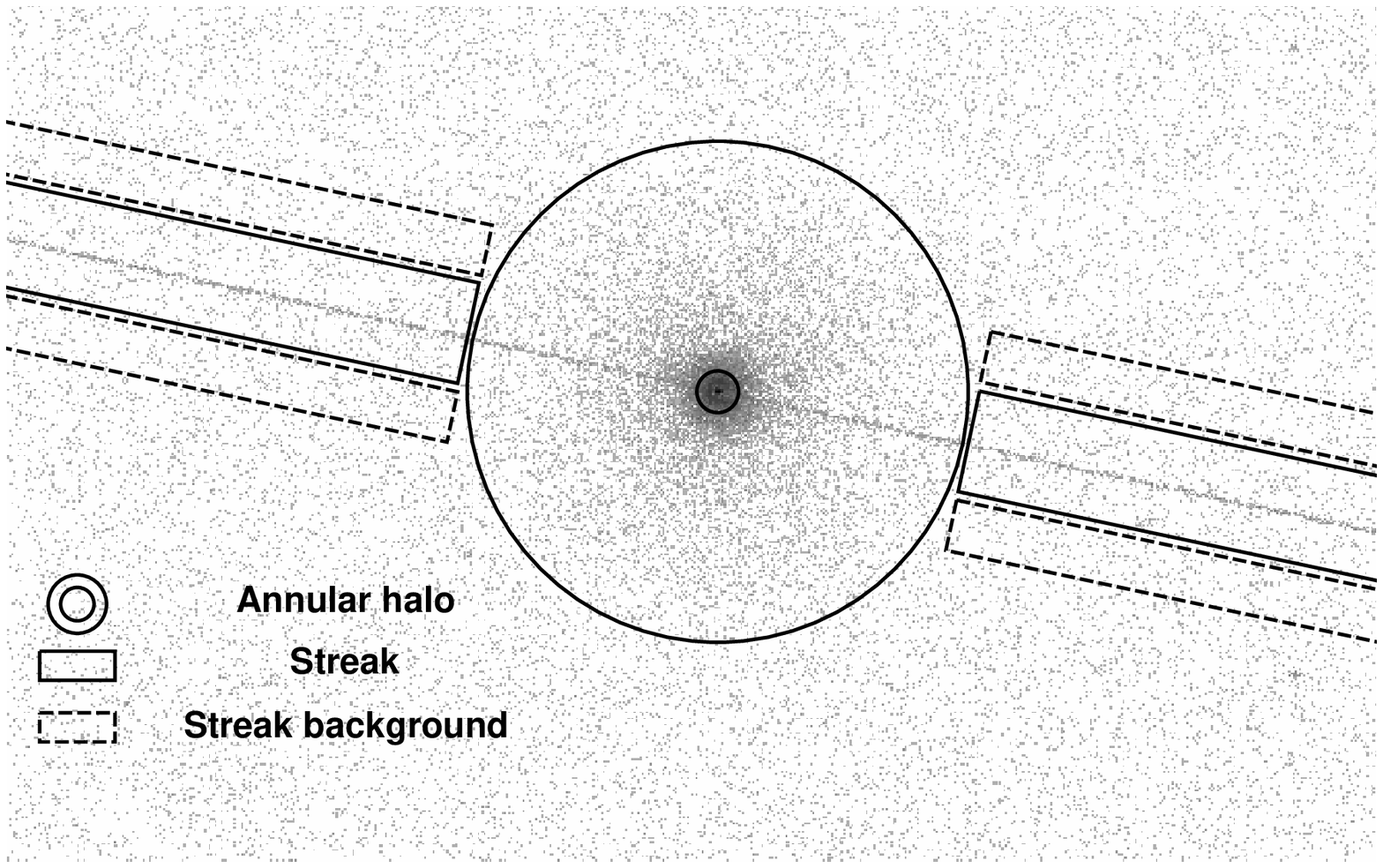}
\caption{2004 July 4 Chandra ACIS--S image (0.5--10~keV) of \src .
The image shows the presence of dust scattered halo
as well as the ACIS readout streak.
The inner radius of the annular halo is 5 arcsec, and
the outer radius of the annular halo is 60 arcsec.
The two pairs of boxes represent the streak area (solid)
and the streak background area (dashed), respectively.
Note that the width of the two pairs of boxes
are enlarged (by a factor of 5) for clarity.
}
\label{fig:saxs_src_img}
\end{figure}

\src\ was observed with ACIS-S onboard \chd\ X--ray Observatory on 2004 July 4 (ObsID 4550).
The detector was operated in time exposure (TE) mode with a time resolution of 3.2 s,
and the total exposure time is 19.06 ks. No grating spectroscopy was used.
The position of \src , ${\rm RA}=17^{\rm h}54^{\rm m}25\fs284$,
${\rm Dec}=-26^\circ 19\arcmin52\farcs62$
($l=3\fdg23$, $b=+0\fdg33$, J2000), was reported by \citet{iza05}.
As shown in Figure~\ref{fig:saxs_src_img}, a diffuse X--ray halo
is present with an extension of $\sim$60 arcsec around the point source.
The data reduction is carried out with CIAO 4.5 and CALDB 4.5.6.

There are two prominent features in Figure~\ref{fig:saxs_src_img}:
the pileup effect and the readout streak.
Pileup\footnote{http://cxc.harvard.edu/ciao/ahelp/acis\_pileup.html}
means that within a single frame (typically, 3.2 s),
at least two events occur
at the same 3 pixel $\times$ 3 pixel island.
The detected energy of these pileup events is approximately
the sum of the individual ones.
If the summed energy exceeds the onboard spacecraft threshold (i.e. 15~keV),
it is rejected automatically by the built-in software of the spacecraft,
probably leading to a visible ``hole" in the image.
\textbf{In other cases, events can be} so close to each other
that their charge clouds overlap significantly,
resulting in grade migration.
Grade migration tends to spread charge into more than one pixel,
degrading the quality of the event
(Gaetz 2010)\footnote{Gaetz, T. 2010, Analysis of the Chandra On-Orbit PSF Wings,
http://cxc.harvard.edu/cal/Acis/detailed\_info.html},
i.e., events with ``good" {\it ASCA} grades (grade 0, 2, 3, 4, 6)
might be converted into ``bad" grades (grade 1, 5, 7).
The presence of the readout
streak\footnote{http://cxc.harvard.edu/ciao/threads/streakextract/}
is due to the fact that the \chd\ ACIS detector system is shutterless.
Hence, photons from the bright source can be detected
while data in the CCD are being read out;
thus the recorded events could have the same CHIPX
as the bright point source, yet locate at any valid CHIPY.

\subsection{Timing analysis}
\label{sct:timing}
\begin{figure}
\centering
\includegraphics[width=0.7\columnwidth,angle=0]{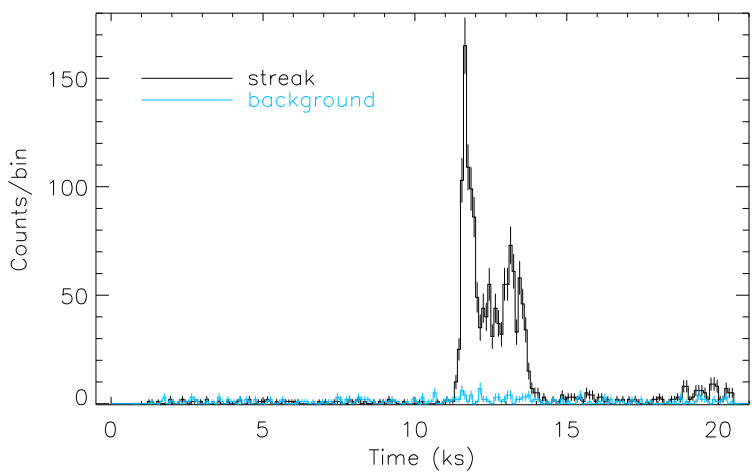}
\caption{The 1--3 keV background subtracted lightcurve of \src\ (ObsID 4550).
The lightcurve has been corrected with proper exposure time
and the time bin is set to 100 s.
The background level (blue), also corrected with proper exposure time,
is presented as well.
}
\label{fig:src_ltc}
\end{figure}

Due to the pileup effect, especially during the outburst,
we extract the lightcurve of \src\ from the ACIS readout streak.
Throughout this work
we set the energy band to be $E\in(1, 3)$ keV unless otherwise stated.
The lower limit is chosen to be 1 keV
because of the poor statistics of X--ray photons below 1 keV,
and the upper limit is set to 3 keV
due to the fact that the contribution of the dust scattered photons
with higher energies are negligible,
as the fractional halo intensity (relative to the source flux)
is proportional to $E^{-2}_{\rm keV}$ \citep{smi02}.
In addition, since the lightcurve is produced from streak data rather than on-axis data,
correction of exposure time should be taken into consideration.
The effective exposure time for the streak area ($T_{\rm str,exp}$) is
\begin{equation}
\small
T_{\rm str,exp}=\frac{T_{\rm exp}}{T_{\rm frm}}\times N \times 0.00004~{\rm s},
\label{str_exp}
\end{equation}
where $T_{\rm exp}$ is the total exposure time of the observation,
$T_{\rm frm}$ is the frame time, {\it N} is the number of rows in the streak area.
After the correction of exposure time,
the 1--3 keV background subtracted lightcurve of \src\
is shown in Figure~\ref{fig:src_ltc}.
By setting a critical count rate of 0.1~cts s$^{-1}$,
we divide the entire observation into the following three stages.
The binary system is in the quiescent stage
for the first $\sim11.0$~ks
with court rate $<$~0.1~cts s$^{-1}$ (denoted as the pre-flare stage),
then a strong flare occurs with a duration of $\sim2.5$~ks (flare stage),
and eventually it returns to the quiescent stage (post-flare stage).

\begin{figure}
\centering
\includegraphics[width=20cm,angle=-90]{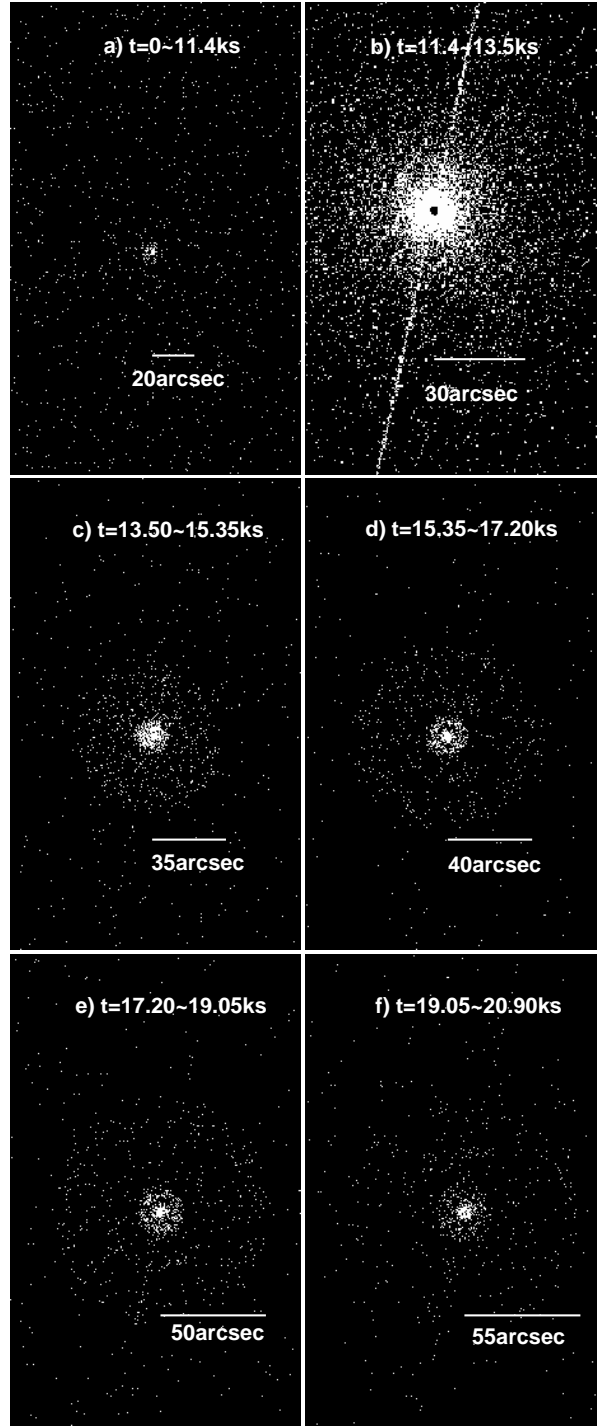}
\caption{The time-dependent image of the expanding \xshalo\ around \src .
Panel a) the image of pre-flare stage with $t\in(0, 11.4)$ ks;
Panel b) the image of flare stage with $t\in(11.4, 13.5)$ ks;
Panel c-f) the images of post-flare stage with $t\in(11.3, 20.9)$ ks.
}
\label{fig:dyn_img}
\end{figure}

The lightcurve is consistent with the time-dependent images of
\src\ (Figure~\ref{fig:dyn_img}). The expanding \xshalo\ around \src\ is
similar to those evolving \xsring s around \textbf{GRBs \citep{vau04, vau06, tie06, via07}
and magnetar bursts \citep{tie10}}.
However, we need to point out that the stacked images
suffer from the contamination of the point spread function (PSF).

\subsection{Imaging Analysis}
\label{sct:imaging}
\subsubsection{Pileup estimation}
\begin{figure}
\centering
\includegraphics[width=0.7\columnwidth,angle=0]{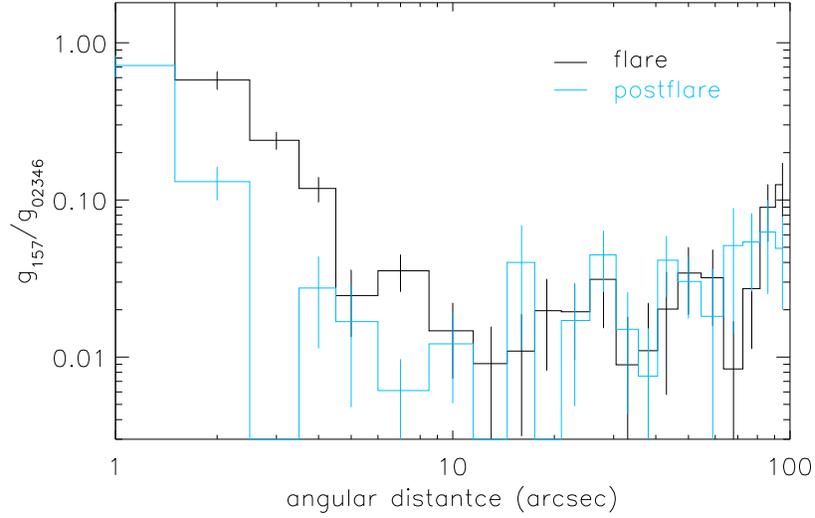}
\caption{The 1--3~keV {\it ASCA} ``bad/good" ratio,
which can be served as a diagnostic of pileup effect.
The black histogram is for the flare stage,
while the blue one is for the post-flare stage.
}
\label{fig:grd}
\end{figure}

\begin{figure}
\centering
\includegraphics[width=0.7\columnwidth,angle=0]{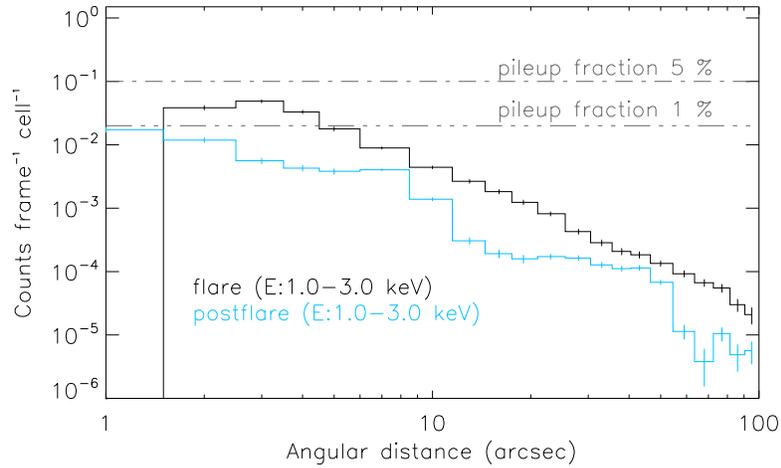}
\caption{The 1--3~keV counts frame$^{-1}$ cell$^{-1}$ ratio
as a function of angular distance for flare and post-flare stages,
which can be served as a diagnostic of pileup effect.
The horizontal dotted-dashed lines indicate
a pileup fraction of 1\% and 5\%, respectively.
}
\label{fig:sfb}
\end{figure}

The interstellar dust in the vicinity of the point source,
if any, might scatter the X--ray photons into
small observation angles ($\lesssim10$ arcsec).
Therefore, we ought to estimate the pileup effect
in order to acquire as much information as possible.
The \src\ ObsID 4550 images the source on
the back-illuminated (BI) chip ACIS-S3,
for which, the $g_{0}/g_{6}$ criterion in estimating the pileup effect
is not as effective as for the front-illuminated (FI) chips,
because the background makes a significant contribution
(Gaetz 2010)\footnotemark[2].
The ``bad/good" ratio (Figure~\ref{fig:grd}) in Level 1 event file
can serve as a pileup indicator;
the gradual rise of the ``bad/good" ratio beyond $\theta \gtrsim 10~{\rm arcsec}$
is due the increasing importance of the background events
with increasing radius (Gaetz 2010)\footnotemark[2].
Meanwhile, using the 3.2 s ACIS frame time and a Poisson-distributed count rate,
we estimate the pileup effect via the same method adopted
in \citet{smi02} and \citet{mcc13}, i.e.
a plot of counts~frame$^{-1}$~cell$^{-1}$ as a function of
radial distance for the flare and post-flare stage.
The pre-flare stage is pileup free with the $E\in(1.0,3.0)$~keV count rate
within a 2.5-arcsec-radius circle centered on the point source
only $\sim4.3\times10^{-3}~{\rm cts}~{\rm s}^{-1}$.
According to Davis (2007)
\footnote{http://cxc.cfa.harvard.edu/csc/memos/ﬁles/Davis\_pileup.pdf},
we take the counts~frame$^{-1}$~cell$^{-1}$ values
for which one would expect pileup fraction of
1\% and 5\% in Figure~\ref{fig:sfb}.
According to both Figure~\ref{fig:grd} and Figure~\ref{fig:sfb},
for $\theta_{\rm obs}\in(4,60)$ arcsec,
the pileup ($\lesssim1\%$) barely affects the observation.
Therefore, we can safely draw the inner boundary of the annular halo,
i.e. the innermost 4.5 arcsec circular region is excluded in the following analysis.
We set two groups of annular halos with different widths depending on the surface brightness:
1) the inner ones are 4.5--6.5 arcsec, 6.5--9.5 arcsec, and 9.5--12.5 arcsec;
2) the outer ones share the same width of 5 arcsec,
with the median angular distance ranging from 15 arcsec to 60 arcsec.

\subsubsection{The Radial profile}
The radial profile of the Level 2 event file is created with the following main steps.
\begin{enumerate}
\item Generate the exposure map (in units of ${\rm ph}^{-1}~{\rm cts~s~cm^{2}}$)
with the CIAO tool
MKEXPMAP\footnote{http://cxc.harvard.edu/ciao/threads/expmap{\_}acis{\_}single/}.
Note that this exposure has taken the quantum efficiency (QE)
and the effective area (ARF) into consideration.
\item Normalize the image (in units of ${\rm cts~pix^{-1}}$)
by the exposure map with the CIAO tool DMIMGCALC\footnotemark[5].
Now the obtained flux image ($F_{\rm img}$)
is in units of ${\rm ph~s^{-1}~cm^{-2}~pix^{-1}}$.
\item Normalize the flux image with the source photon flux ($F_{\rm src}$,
in units of ${\rm ph~s^{-1}~cm^{-2}}$) as follows:
\begin{equation}
\small
P=\frac{\sum_{A} F_{\rm img}}{F_{\rm src}\times A}~{\rm arcsec}^{-2}~,
\label{str_exp}
\end{equation}
where $A$ is the area (in units of ${\rm arcsec}^{2}$) of an annulus
centered at the point source.
\end{enumerate}

\begin{figure}
\centering
\includegraphics[width=0.7\columnwidth,angle=0]{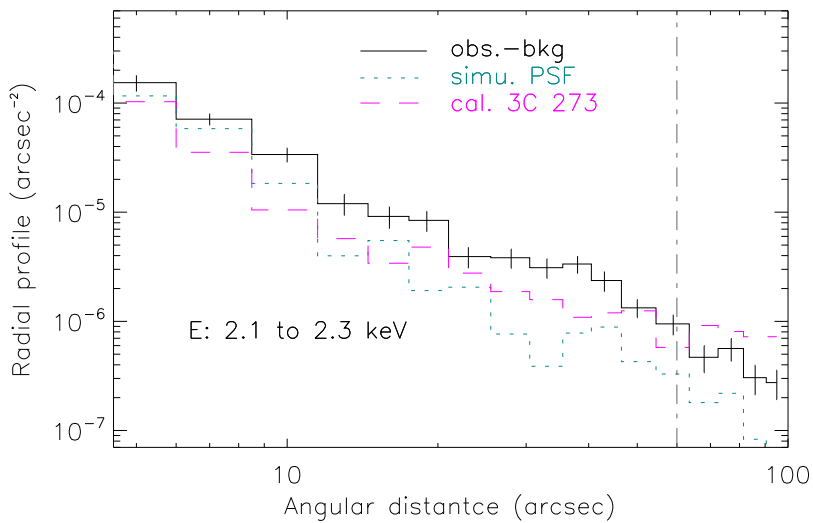}
\caption{The difference between
the background subtracted observational radial profile
(black solid line) and simulated PSF radial profile (cyan dashed line)
shows the existence of the X--ray scattered halo.
The vertical dot-dashed line indicates that for $\theta \gtrsim 60~{\rm arcsec}$,
the simulated PSF underestimates the wing of the genuine PSF.
A background subtracted observational radial profile
of a calibration observation (toward 3C 273)
is used for the PSF radial profile
at $\theta \gtrsim 60~{\rm arcsec}$ instead.
}
\label{fig:hp}
\end{figure}

Likewise, the radial profile of the PSF can also be obtained.
The PSF event file could be simulated with
ChaRT\footnote{http://cxc.harvard.edu/chart/runchart.html}
and MARX\footnote{http://cxc.harvard.edu/chart/threads/marx/},
while the exposure map and the photon flux are the same
as those for the Level 2 event file.
The difference between the background subtracted observational radial profile
and the PSF radial profile shows the existence of the X--ray scattered halo
(Figure~\ref{fig:hp}).
As pointed out by \citet{smi02},
the simulated PSF underestimates the wing of the genuine PSF.
Therefore, a background subtracted observational radial profile of
a calibration observation toward 3C 273 (ObsID 14455) is used
at $\theta \gtrsim 60~{\rm arcsec}$ instead.
Since 3C 273 is located at high galactic latitude ($b=+64\fdg36$, J2000)
and has a low LOS hydrogen column density of $\sim 1.7\times 10^{20}~{\rm cm}^{-2}$,
we believe the contribution of X-ray scattered halo in the radial profile is negligible.

We should be aware that, in our case, the radial profile in Figure~\ref{fig:hp}
underestimates the contribution of the dust scattered halo for two main reasons.
(1) Due to the transient nature of \src,
the dust scattered halo photons are not always there;
however, the exposure time for the entire observation is used in the denominator,
since we do not know the exact exposure time
for the halo photons at a certain angular distance.
(2) The photon flux of \src\ is obtained by
fitting the spectrum with the absorbed power law model
for the entire observation;
however, the photon flux of the flare stage is about three orders of magnitude
greater than that of the quiescent stage (see Table~1 in \citet{ram09}).
Unfortunately, the effective area of \chd\ is so small that
we fail to have sufficient statistics in the quiescent stage
for detailed spectral analysis.

Therefore, the overestimation of both the exposure time
and the source photon flux lead to the underestimation
of the contribution of the dust scattered halo in the radial profile.
Similarly, we also cannot calculate the fractional halo intensity
(FHI; see the definition in \citet{mat91,xia05}),
since the halo is time-dependent.
Thus the definition of FHI works well for persistent systems,
but not very meaningful in terms of the dust scattered halo caused by prompt emission
(e.g. flares of the SFXTs or GRBs).

\section{Distance Measurement}
\label{sct:distance}
\subsection{Cross correlation function}
\label{sct:ccf}
\begin{figure}
\centering
\includegraphics[width=0.7\columnwidth,angle=0]{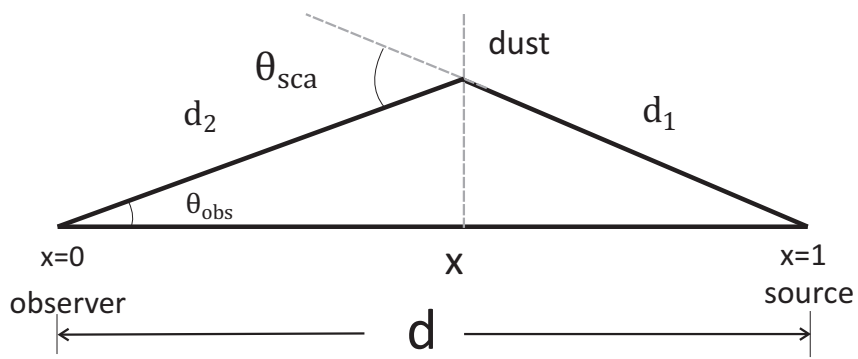}
\caption{A sketch of small angle X--ray scattering. }
\label{fig:saxs_sketch}
\end{figure}

\begin{figure}
\centering
\includegraphics[width=0.7\columnwidth,angle=0]{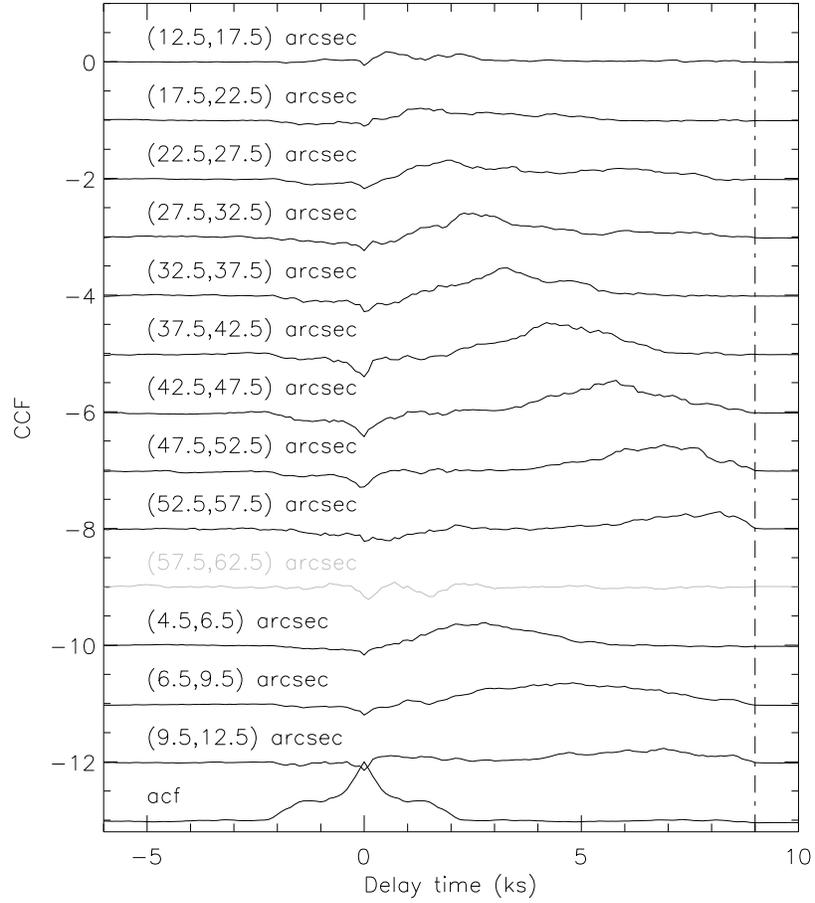}
\caption{The Auto correlation function subtracted cross correlation functions
between the streak lightcurve
and each of the background subtracted observed halo lightcurve.
The vertical dot--dashed line indicate the end of the observation.
For clarity, all but the first CCFs have been lowered by 1.0 successively.
}
\label{fig:ccf}
\end{figure}

As shown in Figure~\ref{fig:saxs_sketch},
the scattered photons travel longer paths than the unscattered ones ($d_1+d_2>d$),
hence it is reasonable to expect time lags of the flare arrival time
in the dust scattered halo lightcurves.
Here we use the cross correlation method \citep{lzx09}
to determine the delay times of the scattered halo photons.
We cross correlate the background subtracted streak lightcurve
with each background subtracted annular halo lightcurve.
The cross correlation function (CCF) is given as follows:
\begin{equation}
\small
{c}(\tau)=\frac{1}{N-|\tau|}\sum_{t=0}^{N-|\tau|-1}\frac{({L}_{\rm h}({t}+\tau)-\overline{L}_{\rm h})}{\sigma_{\rm h}}\frac{({L}_{\rm s}({t})-\overline{L}_{\rm s})}{\sigma_{\rm s}},
\label{eq_ccf}
\end{equation}
where ${c}(\tau)$ is the cross correlation coefficient,
$\tau$ is the delay time, $N$ is the total number of time bins,
$L_X$ is the lightcurve of annular halo ($X=h$) or streak ($X=s$),
$\overline{L}_X$ and $\sigma_X$ are the corresponding mean value
and standard deviation, respectively.
Subsequently, we subtract the auto correlation function (ACF)
of the streak lightcurve, and show the results in Figure~\ref{fig:ccf}.
We conservatively subtract the ACF of the streak lightcurve
rather than subtract the contamination of the PSF contribution
in each annular halo lightcurve, mainly because
there are some uncertainties in the estimation of the PSF contribution.
For instance, as shown in the 2.1--2.3~keV halo profile (Figure~\ref{fig:hp}),
the simulated PSF under-estimates the genuine one at larger angular distances,
so that the PSF fractions obtained from the PSF event file could be biased.
On the other hand, the count rate of the point source
estimated via the count rate of the streak area
also has uncertainty.
Apparently, there are two distinct groups of CCFs shown in Figure~\ref{fig:ccf}.
For annular halos with $\theta_{\rm obs} \in (12.5, 57.5)$ arcsec,
the peaks of CCFs show a clear trend of a shift to the right.
Unfortunately, the peak of the CCF of the halo at $\theta_{\rm obs}=60$ arcsec,
if any, moves out of the end of the this observation
(vertical dot--dashed line at the right in Figure~\ref{fig:ccf}).
On the other hand, the CCFs for the inner three halos
with $\theta_{\rm obs}\in(4.5, 6.5)$ arcsec, $\theta_{\rm obs}\in(6.5, 9.5)$ arcsec
and $\theta_{\rm obs}\in(9.0, 12.5)$ arcsec present relatively longer delay times.

Since the errors of the CCFs obtained above are unavailable analytically,
we turn to Monte Carlo simulations.
Sampled photon counts in each time bin of the lightcurves is generated
either from normal distributions with the mean values set to net photon counts
or from Poisson distributions with the values of $\lambda-$parameters
equal to the net photon counts.
Note that for the majority of those bins which contribute mostly to the peaks of the CCFs,
sufficient counts ($\gtrsim10$) are guaranteed
as we choose the width of each annuli to meet such kind of requirement.
Hence it is reasonable to simply adopt normal distributions here,
although the errors given by normal distributions are smaller
than that of Poisson distributions.
In terms of locating the peaks of the CCFs, again we use two different methods.
One is to fit the $\pm23$ data points centered at the peak of each CCF
with an individual Gaussian function during each realization,
and then determine the mean time delay and the standard deviation
for the time lags for each annular halo.
Alternatively, we simply find the point which yields the maximum value
of the CCF during each realization,
and then calculate the mean values and the standard deviations.
The mean time lags and errors obtained after $10^3$ Monte Carlo realisations
are reported in Table~\ref{tbl:time_lag}.
Note that the uncertainty in this work for each parameter
is given at 68.3\% confidence level, unless otherwise indicated.
Apparently, both the peak values and the numbers of invalid Gaussian fit
suggest that the time lags of the three annular halos (12.5--17.5~arcsec,
17.5--22.5~arcsec and 22.5--27.5~arcsec) are less reliable.
However, the time lags yielded by
the four data sets (Norm./Poi.+Gau./Max.)
of the remaining five annular halos
with $\theta\in(27.5,57.5)$~arcsec agree
within $1\sigma$ uncertainty level.

\subsection{$\theta_{\rm ari}$ and $\theta_{\rm ave}$}
\label{sct:theta}
In practice,
we extract halo lightcurves from wide concentric annuli
around the point source in order to have sufficient counts,
and simply assign the median angular distance
$\theta_{\rm mid}=(\theta_{\rm obs,e}+\theta_{\rm obs,i})/2$
as the $\theta_{\rm obs}$ for each annulus,
\textbf{where $\theta_{\rm obs,e}$ and $\theta_{\rm obs,i}$
are the exterior and interior boundaries of annular regions, respectively}.
However, $\theta_{\rm mid}$ could deviate from $\theta_{\rm obs}$ significantly,
due to sharply and nonlinearly decreasing scattering cross section
as a function of $\theta_{\rm obs}$.
Particularly, for those halos caused by the dust
located in the vicinity of the point source,
the differential scattering cross section
${d\sigma_{\rm sca}}/{d\Omega}$
declines non-linearly
with the increasing scattering angle $\theta_{\rm sca}$,
where $\theta_{\rm sca}=\theta_{\rm obs}/(1-x)$ \citep{mat91}
holds for small observational angles.
The deviation in delay time caused by
$\Delta\theta=\theta_{\rm obs}-\theta_{\rm mid}$ is
\begin{equation}
\Delta t_{\rm dly}=2.42\times 10^{-3} \left(\frac{x}{1-x}\right) \left(\frac{d}{\rm kpc}\right) \left(\frac{\theta \Delta \theta}{{\rm arcsec}^2}\right)~{\rm ks}.
\end{equation}
For instance, assuming $\theta_{\rm obs}=10$~arcsec,
$\Delta \theta_{\rm obs}=1$~arcsec and $d=4$~kpc,
for $x\leqslant0.500$, we have $\frac{x}{1-x}\leqslant1.0$,
and thus $\Delta t_{\rm dly}\leqslant 0.1$~ks.
For $x\geqslant0.909$, however, $\frac{x}{1-x}\geqslant10.0$,
and thus $\Delta t_{\rm dly}\geqslant 1$~ks,
which is comparable to the observed total delay time.
Therefore, when the dust slab is in the vicinity of the point source,
it is important to model $\theta_{\rm obs}$ properly
when determining the delay times of the scattered halo photons.

We extract the observed halo photons \textbf{with $E\in(2.0, 3.0)$ keV} and
$t_{\rm arr}=t_{\rm pk}+t_{\rm dly}\pm t_{\rm dly,err}$
for the three annular regions,
where $t_{\rm arr}$ is the arrival time of the halo photons,
$t_{\rm pk}$ is the time when the streak lightcurve
(used as a proxy for the source lightcurve) reaches its maximum,
$t_{\rm dly}$ and $t_{\rm dly,err}$ are listed in Table \ref{tbl:time_lag}.
The time intervals are set so that
the dust scattered halo photons (net counts)
dominates the PSF photons and the background photons out there.
In fact, the background counts ($\lesssim 10^{-1}$) are negligible here.
While for the PSF photons, we run ChaRT and MARX to simulate the observation
and generate $\sim10^3$ PSF photons within each of the three annuli.
We subtract the contribution from the PSF photons,
and then list in Table~\ref{tbl:ari_df}
the arithmetic mean angular distance of each annulus,
\begin{equation}
\theta_{\rm ari}=\frac{1}{N}~\sum_{i} \theta_i,
\end{equation}
where $\theta_i$ is the angular distance of the $i$th photon
and $N$ is the total number of photons in this annulus, respectively.
Apparently, when the dust slab is close to the point source,
$\theta_{\rm ari}$ does differ from $\theta_{\rm mid}$.

To be more specific, we can calculate the single-scattering cross section
with Rayleigh-Gans (RG) approximation of
the differential scattering cross section \citep{mat91}
\begin{eqnarray}
\frac{d\sigma_{\rm sca}}{d\Omega} &=& c_1 \left(\frac{2Z}{M}\right)^2 \left(\frac{\rho}{{\rm 3\ g\ cm}^{-3}}\right)^2 \left(\frac{a}{\mu\rm{\ m}}\right)^6 \nonumber \\
&&\times\left[\frac{F(E)}{Z}\right]^2 {\rm exp}\left(-K^2\left(\frac{\theta_{\rm sca}}{\rm arcmin}\right)^2\right)~,
\end{eqnarray}
where $c_1=9.31\times10^{-8}~{\rm cm}^2{\rm arcmin}^{-2}$,
$Z$ is the mean atomic charge, $M$ is the molecular weight (in units of amu),
$\rho$ is the mass density,
$F(E)$ is the atomic scattering factor,
and $K^2=0.4575 (E/{\rm keV})^2(a/{\mu{\rm m}})^2$.
As pointed out by \citet{smi98}, the RG approximation fails for
incident photons with energies $E\lesssim2$~keV.
Hence, in the following analysis
we only focus on photons with $E\gtrsim2$~keV.
With one thin dust slab located at $x=x_i$ along the LOS,
the single-scattering cross section at $\theta_{\rm obs}$ is
\begin{eqnarray}
\sigma_{\rm sca}(x=x_i,\theta_{\rm obs}) &=& \int S(E)dE \int f(x_i) N_{\rm H} n(a) \nonumber \\
&&\times \frac{d\sigma_{\rm sca}}{d\Omega}(a,E,\theta_{\rm obs},x_i) da~,
\label{eq:crs}
\end{eqnarray}
where $S(E)$ is the normalized photon energy distribution,
$n(a)$ is the dust size distribution
(in units of particles per H atom per micron),
$f(x_i)$ is the density of hydrogen at $x\cdot d$
relative to the average hydrogen column density along the LOS to \src\
and here we set $f(x_i)$ to unity.
For simplicity, Equation~\ref{eq:crs} is substituted with
\begin{eqnarray}
\sigma_{\rm sca}(x=x_i,\theta_{obs}) &=& \sum_{k=0}^{9}n(E_k) \int f(x_i) N_{\rm H} n(a) \nonumber \\
&&\times \frac{d\sigma_{\rm sca}}{d\Omega}(E_i,a,\theta_{\rm obs},x_i) da~, \nonumber
\end{eqnarray}
where $n(E_k)$ is the normalized observed spectrum
within the range of $E\in(E_k-0.05,E_k+0.05)$~keV,
$E_k=(2.05+0.1k)$~keV, and $\sum n(E_k)=1$.
We obtain the average observational angles ($\theta_{\rm ave}$)
predicted by MRN, WD01, ZDA and XLNW dust models via
\begin{equation}
\theta_{\rm ave}=\frac{\int \frac{d\sigma_{\rm sca}}{d\Omega}(x=x_i,\theta_{\rm obs})\theta_{\rm obs}^{2}d\theta_{\rm obs}}{\int \frac{d\sigma_{\rm sca}}{d\Omega}(x=x_i,\theta_{\rm obs})\theta_{\rm obs}d\theta_{\rm obs}}~.
\label{eq:ave_theta}
\end{equation}
An advantage of Equation~\ref{eq:ave_theta} is that
$\theta_{\rm ave}$ does not depend on the distance of the point source.
Consequently, we can break the degeneracy between the distances of the point source
and the dust slab in Equation~\ref{eq:time_lag}.

\subsection{Distance measurement}
\label{sct:dist}
The relationship between the delay time and geometrical distances of interstellar dust
and the point source is given by \citet{tru73},
\begin{equation}
\left(\frac{t_{\rm dly}}{\rm ks}\right)=1.21 \times 10^{-3} \frac{x}{1-x} \left(\frac{d}{\rm kpc}\right) \left(\frac{\theta_{\rm obs}}{\rm arcsec}\right)^2,
\label{eq:time_lag}
\end{equation}
where $x$ is the normalized distance of the dust cloud,
$d$ is the distance of the point source,
$\theta_{\rm obs}$ is the observational angle of the halo photons
and here we simply \textbf{assign $\theta_{\rm obs}=\theta_{\rm mid}$}.
In fact, \citet{rah08} reported a distance of 3.6 kpc
for \src\ using mid-infrared photometry and spectroscopy.
The result is within the range $d\in(2.1, 4.2)$ kpc given by \citet{pel06}.

We first adopt a distance of 3.6 kpc for \src\
and fit the delay times for the halos caused by the closer dust
to determine the normalized dust distance ($x$);
the results are listed in Table~\ref{tbl:x_d_fixed}.
Note that for those fits with reduced chi-squared
values greater than 3, we only report values of $x$
yielding the smallest reduced chi-squared values.
The results obtained from the four sets of data
are consistent with each other within the 68.3\% confidence level.
In terms of the halos caused by the farther dust,
the reduced chi-squared values are significantly greater than unity,
so that the normalized distances of the farther dust cloud are less reliable.
Simply by solving Equation~(\ref{eq:time_lag}) with the three time lags
of the Poi.+Max. data, we have three normalized dust distances,
$0.952^{+0.004}_{-0.005}$ for $\theta_{\rm obs}\in(4.5, 6.5)$~arcsec,
$0.944^{+0.004}_{-0.004}$ for $\theta_{\rm obs}\in(6.5, 9.5)$~arcsec
and $0.925^{+0.006}_{-0.008}$ for $\theta_{\rm obs}\in(9.5, 12.5)$~arcsec,
which indicates that the farther dust could
probably be a complex or a giant molecular cloud;
alternatively, the simple model needs to be modified.

Since no uncertainty of the point source distance
was reported in \citet{rah08},
here we attempt to do distance measurement
via the X-ray scattered halo.
Since the two parameters $x$ and $d$
in Equation~(\ref{eq:time_lag}) are highly degenerated and negatively correlated,
we introduce a parameter called distance factor
\begin{equation}
\mathscr{D}=\frac{x}{1-x}\frac{d}{\rm kpc},
\label{eq:df}
\end{equation}
which contains information of dust distance and source distance
and can be determined with the delay time of the scattered halo photons.
Substituting $\mathscr{D}$ into Equation~(\ref{eq:time_lag}), we have
\begin{equation}
\left(\frac{t_{\rm dly}}{\rm ks}\right)=1.21 \times 10^{-3} \mathscr{D} \left(\frac{\theta_{\rm obs}}{\rm arcsec}\right)^2~.
\label{eq:time_lag_df}
\end{equation}
The results for $\mathscr{D}$ are listed in Table~\ref{tbl:df}.
Apparently, for the closer dust cloud,
$\mathscr{D}$ could be well constrained
for the \textbf{Norm./Poi.+Max.} data sets;
whereas for the farther dust cloud,
the values of $\mathscr{D}$ fail to agree for all four data sets.
Moreover, using $\theta_{\rm ari}$ instead of $\theta_{\rm mid}$
when fitting data with Equation~(\ref{eq:time_lag_df})
cannot eliminate the discrepancy.
The inconsistency is mainly due to the fact that
even a small deviation in $x$ ($\Delta x\sim0.01$) could lead to
a substantial change in $\mathscr{D}$
with $\Delta \mathscr{D}\sim 10$.

In order to determine the point source distance ($d$),
we need to firstly obtain the normalized distance of the dust ($x$)
by comparing the arithmetic mean values
($\theta_{\rm ari}$) of the observed scattered halo photons
within different annular regions with the average mean values
($\theta_{\rm ave}$) calculated with dust grain models.
Meanwhile, the time lags of the scattered halo photons
obtained through the cross correlation method
and $\theta_{\rm ari}$ allow us to determine
 $\mathscr{D}$,
which is a function of both $d$ and $x$.
Substituting $x$ into Equation~(\ref{eq:df}),
$d$ is derived finally.

\begin{figure}[h]
\centering
\includegraphics[width=0.7\columnwidth,angle=0]{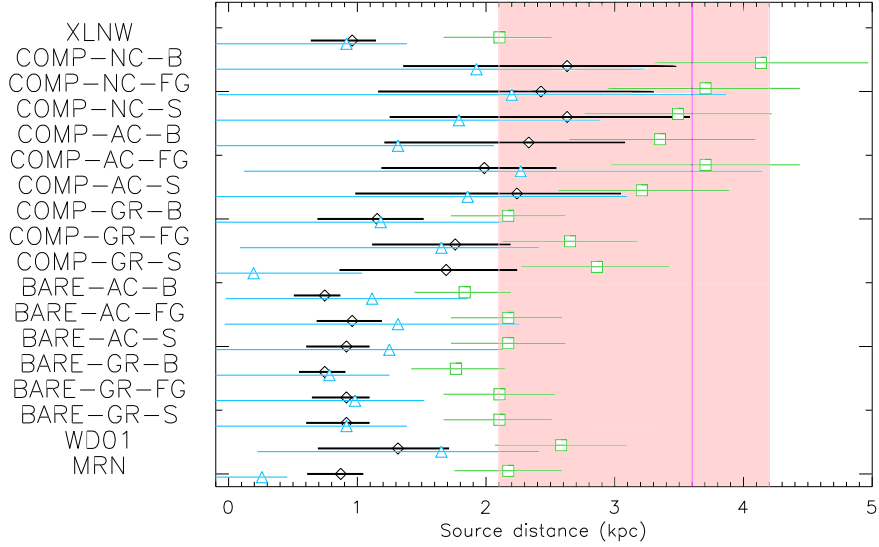}
\caption{The distances of \src\ determined with different dust grain models.
The black diamond, blue triangles and green squares
indicate the results obtained with $\theta\in(27.5, 52.5)$~arcsec,
$\theta\in(6.5, 9.5)$~arcsec and $\theta\in(4.5, 12.5)$~arcsec
annular halos, respectively.
The pink region and solid magenta line show
the results of the two IR observations.
}
\label{fig:d_src}
\end{figure}

To be more specific, for the inner three individual annular halos
(4.5--6.5~arcsec, 6.5--9.5~arcsec
and 9.5--12.5~arcsec) caused by the farther dust,
by varying $x\in(0.900, 0.999)$
with step $\Delta x=0.001$,
we find the minimum values of $|\theta_{\rm ave}-\theta_{\rm ari}|$
(for $E\in(2.0, 3.0)$~keV) and the best $x$.
Unfortunately, due to the low counts,
the uncertainty of $\theta_{\rm ari}$ is rather large.
Consequently, even the derived $x$
of the 6.5--9.5~arcsec annular halo,
which has the smallest uncertainty on $\theta_{\rm ari}$,
can barely constrain $d$
(\textbf{see the blue triangles in Figure~\ref{fig:d_src}}).

Alternatively, we use the combination of
the individual annular halos caused by the farther dust cloud.
Due to the sufficient counts within $\theta\in(4.5, 12.5)$~arcsec
$x$ can be well constrained (Table~\ref{tbl:theta_comb}).
As for $\mathscr{D}$,
we simply \textbf{set},
\begin{equation}
\mathscr{D}=\frac{1}{3}{\sum_{i=0}^{2}\mathscr{D}_{i}}~,
\delta \mathscr{D}=\sqrt{\sum_{i=0}^{2}\delta \mathscr{D}_{i}^{2}}~.
\label{eq:df_comb}
\end{equation}
Due to the large uncertainty in $\mathscr{D}$,
the source distance cannot be narrowly constrained.

On the other hand, in terms of the annular halos
caused by the closer dust cloud,
we combine the four annular halos with $\theta\in(27.5, 52.5)$~arcsec,
and vary $x\in(0.20, 0.90)$ with step $\Delta x=0.01$
to search for the minimum values of $|\theta_{\rm ave}-\theta_{\rm ari}|$
and the best $x$. $\mathscr{D}$ is well constrained
via the cross correlation method for the closer dust. However, the uncertainty in $x$ is quite large
due to the relatively low counts in such a wide region.
Thus, the source distance cannot be well constrained neither
(see also Table~\ref{tbl:theta_comb}).

Not all of the point source distances derived above are reasonable
when compared to the results obtained with  IR observations
\citep{pel06,rah08}.
Figure~\ref{fig:d_src} illustrates the source distances
obtained with different dust grain models.
Given the distance range $d\in(2.1, 4.2)$~kpc \citep{pel06},
it seems that four dust grain models COMP-AC-S/B, COMP-NC-S/FG
(labeled with $\surd$ in Table~\ref{tbl:theta_comb}) are better,
\textbf{since both $d_1$ and $d_2$ are within the distance range,
i.e. $d_i\in (2.1, 4.2)~{\rm kpc},i=1~{\rm and}~2$.}
COMP-GR-S/FG, COMP-AC-FG, COMP-NC-B \textbf{(labeled with $\bigcirc$)} are also acceptable,
\textbf{since either $d_1$ or $d_2$ is within the distance range,
while the other is consistent with the distance range within $1\sigma$ error,
i.e. $|d_{\rm b}-d_i|\in d_{i, \rm err},i=1~{\rm or}~2$,
where the upper and lower boundary of the distance range
$d_{\rm b}=2.1~{\rm and}~4.2$ kpc, respectively.
However, for the rest of the dust grain models,
$d_1$ is within the distance range
or consistent within 1$\sigma$ uncertainties (BARE-GR/AC-B),
while $d_2$ is inconsistent with the distance range within 1$\sigma$ uncertainties
($|2.1-d_2|>d_{2,\rm err}$).}

\section{Discussion}
\label{sct:discussion}
\subsection{Kinematic distance measurements of molecular clouds}
\label{sct:clouds_dist}
For $d\sim3.6$~kpc \citep{rah08},
the distances for the closer and farther dusts are
$\sim1.8$~kpc and $\sim3.4$~kpc away from us, respectively.
In this subsection we try to find kinematic distance measurements
of the molecular clouds along the LOS toward \src\,
for comparison with our geometrical distances of the dust slabs.
A rough estimate of the radial velocity of the molecular clouds
where the dust slab is embedded can be made via \citep{rdu09}
\begin{equation}
r=R_\odot \sin l \frac{V(r)}{v_{\rm los}+V_\odot \sin l},
\label{v_los}
\end{equation}
where $r$ is the distance of the molecular cloud to the Galactic Center (GC),
$R_{\odot}$ is the distance of the Sun to GC,
{\it l} is the galactic longitude of the LOS,
$V_{\odot}$ is the rotation velocity of the Sun,
$V(r)$ is the rotation velocity of the molecular cloud,
and $v_{\rm los}$ is the projection of $V(r)$ to the LOS,
also known as the radial velocity.
Assuming $R_{\odot}=8.5$~kpc, $V_{\odot}=220$~km s$^{-1}$
and a flat rotation curve (i.e. $V(r)=220$~km s$^{-1}$),
we derive $v_{\rm los}$ of $\sim3.1$~km s$^{-1}$
and $\sim7.8$~km s$^{-1}$
for the closer and farther molecular cloud, respectively.
In fact, \citet{dah98} reported an averaged $v_{\rm los}\sim5$~km s$^{-1}$
for $^{12}$CO(1-0) emission line for the Southern Clump 2 region,
$l\in(2\fdg7, 3\fdg5)$, $b\in(0\fdg15, 0\fdg35)$,
which is roughly consistent with our result of the farther dust.

\subsection{Issues with observational angles}
\label{sct:caveats}
In Section~\ref{sct:theta}, we have shown that
given the normalized distance of a dust slab,
different interstellar dust models predict different average observational angles
for scattered photons within certain annular regions.
This offers the advantage of estimating the parameter $x$ from data,
thus breaking the degeneracy between $x$ and $d$.
We tested several interstellar dust grain models with the data of the farther dust,
for which the predictions of these models become quite different,
because the scattering angle $\theta_{\rm sca}=\theta_{\rm obs}/(1-x)$
is quite large for the farther dust.
It is therefore possible to distinguish among different interstellar dust models,
as demonstrated above.
However in the calculations of the scattering differential cross section,
Gaussian approximation is used for the form factor
in the RG approximation.
\citet{smi98} pointed out that the Gaussian approximation leads to deviations
at large scattering angle ($\gtrsim200~{\rm arcsec}$) and large dust grain size.
In our case, i.e., $x>0.95$ and $\theta_{\rm obs}\sim10~{\rm arcsec}$
for the farther dust slab, we have $\theta_{\rm sca}>200$~arcsec.
Moreover, the upper limits of the grain size are greater than 0.3~micron,
and even reach to 0.9~micron.
Therefore the Gaussian approximation may cause considerable inaccuracies
to the model predictions.
Alternatively, one can turn to Mie theory \citep{vhu57},
which is more accurate for $E\sim1~$keV and/or large scattering angles,
but numerically more difficult to carry out the calculations.

\section{Summary}
\label{sct:summary}
In this work, we analyzed the \xshalo\ around \src\ with cross correlation method.
The main results are summarized as follows:
\begin{enumerate}
\item From the cross correlation functions
between the streak lightcurve (used as a proxy for the point source lightcurve)
and the lightcurves of the annular halos, we conclude that
there are at two interstellar dust clouds along the LOS toward \src .

\item By comparing
the observational angle of the scattered halo photons
with that predicted by different dust grain models,
the normalized dust distance can be determined
independent of the distance of the point source.

\item Given the point source distance of $\sim3.6$~kpc,
the closer dust, which is $\sim 1.8$~kpc away from us, is responsible for
\xshalo s at larger observational angles ($\gtrsim12.5$ arcsec).
The farther dust, which is quite close to the point source ($\sim3.4$~kpc away from us),
explains the X--ray scattered halos at smaller angular distances ($\lesssim12.5$ arcsec).

\item We determined the model-dependent point source distances,
which are compared with that yielded by IR observations.
We find that among the 18 tested dust grain models
(MRN, WD01, ZDAs and XLNW),
the four dust grain models COMP-AC-S/B, COMP-NC-S/FG are better,
COMP-GR-S/FG, COMP-AC-FG, COMP-NC-B are also acceptable,
but the rest dust grain models fail to
obtain consistent source distance.
\end{enumerate}

Similar to the GRBs, the transient nature of SFXTs
can, in principle, be used to precisely determine
the geometrical distances of interstellar dust and the point source
by taking advantage of the time delay effect of
the small angle X--ray scattering phenomena.
However, we have to face some practical difficulties, such as:
1) the angular resolutions of the space telescopes are relatively poor;
2) the effective area is small and thus the photon counts are relatively low; or
3) for observations of \xshalo\ caused by dust slab in the vicinity of the point source,
the time lags can be quite large, but
no observations with sufficiently long effective exposure times are available.
The effective exposure time refers to the exposure time for observing the \xshalo .
For instance, in our work,
since the outburst occurs $\sim$~10~ks after the beginning of this observation,
the effective exposure time for observing the \xshalo\ here is only $\sim$~9 ks.

With the fine angular resolution,
\chd\ has the capability to observe \xshalo s around SFXTs
especially at smaller angular distance,
although the collecting area of ACIS is small.
Unfortunately, insofar as the archival \chd\ data,
only \src\ (ObsID 4550) allows us to study the \xshalo .
In terms of other observations of SFXTs,
either the exposure time is only several kiloseconds (e.g. XTE J1739$-302$),
or no flaring activity was caught (e.g., for IGR J19410$-$0951).
Therefore, we suggest that in the future more long term follow up observations
of the outbursts of SFXTs shall be made with \chd\
to study the \xshalo\ and thus the interstellar dust models in further details.

\acknowledgments
SNZ acknowledges partial funding support by 973 Program of China under grant 2009CB824800,
by the National Natural Science Foundation of China under grant Nos. 11133002, 11373036
and 10725313, and by the Qianren start-up grant 292012312D1117210.

\appendix

\section{Thick Dust Layer}
The smallest size of the giant molecular clouds (GMCs)
in the Milky Way has been found to be 5~pc \citep{mur11}.
Therefore the farther dust cloud located $\lesssim100$~pc away
from the binary system is no longer ``thin" when compared to the distance
between the cloud and the point source.
Consequently, the validity of the treatment of the dust cloud
as a ``thin" slab should be examined.
Consider a point source at a distance of 2~kpc
along with a ``thick" dust cloud with a thickness of 10 pc located at$x \gtrsim 0.90$),
the dust scattering cross section of such a cloud can be calculated as
the sum of five ($N=5$) ``thin" slabs
\begin{equation}
\sigma_{\rm thick}=\sum_{i=0}^{4}~\int S(E)dE \int \frac{N_{\rm H}}{N}n(a) \times \frac{d\sigma_{\rm sca}}{d\Omega}(a,E,\theta_{\rm obs},x_i=x_0+i\times \Delta x) da.
\label{eq:crs_thick}
\end{equation}
For simplicity, we assume a mono-energy spectrum ($E=2$~keV) and $\theta_{\rm obs}=10$~arcsec.
We compare $\sigma_{\rm thick}$ with the cross section of
a single ``thin" dust slab located at $x=\bar x_i$,
which has a thickness of 2 pc and the same total column density as that of the ``thick" cloud,
by calculating
\begin{equation}
R=\frac{\frac{1}{N}\sum_{i=0}^{4}\int n(a) \times \frac{d\sigma_{\rm sca}}{d\Omega}(a,E,\theta_{\rm obs},x_i=x_0+i\times \Delta x) da}{\int n(a) \times \frac{d\sigma_{\rm sca}}{d\Omega}(a,E,\theta_{\rm obs},x_i=\bar x) da}
\label{eq:crs_ratio}
\end{equation}
for four typical dust grain models
(MRN, WD01, XLNW and ZDA COMP-GR-S)
with $\bar x=0.900,~0.950~{\rm and}~0.990$ (Table~\ref{tbl:crs_r}), respectively.
Note that the XLNW model is a modified form of ZDA BARE-GR-S model \citep{xia11}.
Clearly in all cases $R$ does not deviates from unity significantly,
indicating that the ``thin" dust cloud assumption is a good approximation
when $x\lesssim 0.990$ and the size of the dust cloud is not significantly large ($\lesssim 10~$pc).

%\bibliographystyle{apj}
%\bibliography{citation}

%%-----------------------------Figure Start------------------------------
%%-----------------------------Figure End------------------------------

%-----------------------------Figure Start------------------------------
%-----------------------------Figure End------------------------------

%-----------------------------Figure Start------------------------------
%-----------------------------Figure End------------------------------

%-----------------------------Figure Start----------------------------
%-----------------------------Figure End------------------------------
%-----------------------------Figure Start------------------------------
%-----------------------------Figure End------------------------------

%-----------------------------Figure Start----------------------------
%-----------------------------Figure End------------------------------

%-----------------------------Figure Start------------------------------
%-----------------------------Figure End--------------------------------

%-----------------------------Figure Start------------------------------
%-----------------------------Figure End------------------------------

%-----------------------------Figure Start------------------------------
%-----------------------------Figure End------------------------------

%-----------------------------Table Start------------------------------
\begin{deluxetable}{ccccccccc}
\tabletypesize{\scriptsize}
%\rotate
\tablecolumns{9}
\tablewidth{0pc}
\tablecaption{Time lags of the annular halos. \label{tbl:time_lag}}
\tablehead{$\theta_{\rm obs}$ & Norm.+Gau. &  & Norm.+Max.  &
& Poi.+Gau. &  & Poi.+Max.   &  \\
arcsec & $t_{\rm dly}$ (ks) & C$_{\rm max}^{\rm a}$ ($n^{\rm b}$) & $t_{\rm dly}$ (ks) & C$_{\rm max}$
& $t_{\rm dly}$ (ks) & C$_{\rm max} (n)$ & $t_{\rm dly}$ (ks) & C$_{\rm max}$}
\startdata
12.5-17.5    &  $2.84\pm0.09$  & 0.00 (997) &  $0.95\pm0.64$ & 0.18 &  $2.72\pm0.92$  & 0.00 (977) &  $1.10\pm0.72$  & 0.17   \\
17.5-22.5    &  $2.88\pm0.37$  & 0.01 (892) &  $1.46\pm0.44$ & 0.22 &  $3.03\pm0.37$  & 0.03 (797) &  $1.59\pm0.58$  & 0.22   \\
22.5-27.5    &  $2.82\pm0.40$  & 0.13 (327) &  $1.99\pm0.53$ & 0.30 &  $2.89\pm0.49$  & 0.16 (175) &  $2.08\pm0.64$  & 0.30   \\
27.5-32.5    &  $3.04\pm0.15$  & 0.24 (209) &  $2.50\pm0.24$ & 0.39 &  $3.02\pm0.13$  & 0.29 (61)  &  $2.48\pm0.20$  & 0.40   \\
32.5-37.5    &  $3.50\pm0.12$  & 0.35 (13)  &  $3.24\pm0.12$ & 0.44 &  $3.50\pm0.10$  & 0.36 (2)   &  $3.23\pm0.11$  & 0.45   \\
37.5-42.5    &  $4.64\pm0.12$  & 0.46 (1)   &  $4.37\pm0.26$ & 0.50 &  $4.66\pm0.10$  & 0.47 (0)   &  $4.37\pm0.25$  & 0.50   \\
42.5-47.5    &  $5.60\pm0.13$  & 0.46 (0)   &  $5.72\pm0.19$ & 0.50 &  $5.53\pm0.11$  & 0.45 (0)   &  $5.70\pm0.19$  & 0.50   \\
47.5-52.5    &  $6.71\pm0.20$  & 0.35 (32)  &  $6.99\pm0.43$ & 0.41 &  $6.64\pm0.17$  & 0.36 (4)   &  $6.88\pm0.40$  & 0.41   \\
52.5-57.5    &  $7.59\pm0.42$  & 0.14 (428) &  $7.99\pm0.37$ & 0.28 &  $7.56\pm0.54$  & 0.23 (35)  &  $7.81\pm0.45$  & 0.27   \\
57.5-62.5    &  $-1.88\pm7.42$ & 0.00 (989) &  $0.90\pm3.51$ & 0.10 &  $-7.27\pm3.66$ & 0.16 (846) &  $-0.55\pm4.16$ & 0.12   \\
\hline
\noalign{\smallskip}
4.5-6.5      &  $3.06\pm0.13$  & 0.33 (43)  &  $2.60\pm0.30$ & 0.38 &  $3.08\pm0.12$  & 0.34 (22)  &  $2.63\pm0.27$  & 0.39   \\
6.5-9.5      &  $4.70\pm0.23$  & 0.35 (1)   &  $4.66\pm0.38$ & 0.35 &  $4.79\pm0.20$  & 0.36 (0)   &  $4.70\pm0.34$  & 0.37   \\
9.5-12.5     &  $6.33\pm0.36$  & 0.20 (30)  &  $6.57\pm0.99$ & 0.23 &  $6.09\pm0.34$  & 0.21 (2)   &  $6.48\pm0.63$  & 0.24   \\
\enddata
\tablecomments{``Norm.": the sampled photon counts generated
from normal distributions; ``Poi.":  the sampled photon counts generated
from Poisson distributions; ``Gau.": an individual Gaussian function is used to fit the nearby data points centered at the peak of each CCF; ``Max.": the maximum value of the CCF is used.}
\tablenotetext{a}{~The peak value of the CCFs.}
\tablenotetext{b}{~The number of bad fits to a Gaussian function.}
\end{deluxetable}
%-----------------------------Table End------------------------------

%-----------------------------Table Start------------------------------
\begin{deluxetable}{cccccc}
\tabletypesize{\footnotesize}
%\rotate
\tablecolumns{6}
\tablewidth{0pc}
\tablecaption{$\theta_{\rm ari}$ and distance factor ($\mathscr{D}$)
for scattered halo photons with $E\in(2.0,3.0)$~keV. \label{tbl:ari_df}}
\tablehead{annuli~(arcsec) & (4.5, 6.5) & (6.5, 9.5) & (9.5, 12.5) & (4.5, 12.5) & (27.5, 52.5)  \\
$t_{\rm arr}~(\rm ks)$ & (14.0, 14.6) & (16.0, 16.7) & (17.5, 18.8) & (12.5, 19.1) & (12.5, 19.1)}
\startdata
number of net counts         & 18.86          & 40.51          & 33.19          & 347.45         & 288.24          \\
$\theta_{\rm ari}$~(arcsec)  & $5.25\pm0.14$  & $7.76\pm0.14$  & $10.63\pm0.17$ & $7.56\pm0.11$  & $39.79\pm0.42$ \\
$\mathscr{D}$                & $78.91\pm9.16 ^{\rm a}$ & $64.67\pm5.18 ^{\rm a}$ & $47.40\pm4.86 ^{\rm a}$ & $63.66\pm11.59 ^{\rm b}$ & $2.24\pm0.07 ^{\rm c}$   \\
\enddata
\tablenotetext{a}{~For the inner three annular regions,
the distance factors ($\mathscr{D}$) are obtained by Equation~(\ref{eq:time_lag_df})
with the Poi.+Max. $t_{\rm dly}$ in Table~\ref{tbl:time_lag}
and $\theta_{\rm ari}$ in this table.}
\tablenotetext{b}{~In this combined annular region,
$\mathscr{D}$ is obtained via Equation~(\ref{eq:df_comb}).}
\tablenotetext{c}{~Here, $\mathscr{D}$ is the taken from the Poi.+Max.
$\mathscr{D}$ in Table~\ref{tbl:df}.}
\end{deluxetable}
%-----------------------------Table End------------------------------

%-----------------------------Table Start------------------------------
\begin{deluxetable}{ccccc}
\tabletypesize{\footnotesize}
%\rotate
\tablecolumns{5}
\tablewidth{0pc}
\tablecaption{The normalized distances ($x$) of dust
with source distance ($d$) fixed to 3.6~kpc.\label{tbl:x_d_fixed}}
\tablehead{ & Norm.+Gau. & Norm.+Max. & Poi.+Gau. & Poi.+Max.}
\startdata
$\chi^2$/d.o.f.             &  20.9/5           &  3.7/8            &  27.5/5   &  4.5/8                         \\
$x$ for (12.5, 57.5)~arcsec &  $0.52$           &  $0.518\pm0.005$  &  $0.52$   &  $0.517\pm0.005$               \\
$\chi^2$/d.o.f.             &  88.8/2           &  6.5/2            &  115.8/2  &  13.1/2                        \\
$x$ for (4.5, 12.5)~arcsec  &  $0.96$           &  $0.96$           &  $0.96$   &  $0.96$                        \\
\enddata
\end{deluxetable}
%-----------------------------Table End------------------------------

%-----------------------------Table Start------------------------------
\begin{deluxetable}{cccccc}
\tabletypesize{\footnotesize}
%\rotate
\tablecolumns{6}
\tablewidth{0pc}
\tablecaption{Determining the distance \textbf{factor} $\mathscr{D}$
by fitting the time lags with Equation~(\ref{eq:time_lag_df}).\label{tbl:df}}
\tablehead{$\theta_{\rm obs}$~(arcsec) & para. & Norm.+Gau. & Norm.+Max. & Poi.+Gau. & Poi.+Max.}
\startdata
(12.5, 57.5)            & $\chi^2$/d.o.f. & 20.9/5 & 3.7/8         & 27.5/5  & 4.5/8          \\
with $\theta_{\rm mid}$ & $\mathscr{D}$   & $2.3$  & $2.26\pm0.04$ & $2.3$   & $2.24\pm0.07$   \\
\tableline
\noalign{\smallskip}
(4.5, 12.5)             & $\chi^2$/d.o.f. & 88.8/2 & 6.5/2         & 115.7/2 & 13.1/2         \\
with $\theta_{\rm mid}$ & $\mathscr{D}$   & $57.7$ & $58.0$        & $57.9$  & $55.2$          \\
\tableline
\noalign{\smallskip}
(4.5, 12.5)             & $\chi^2$/d.o.f. & 91.6/2 & 6.6/2         & 118.6/2 & 13.0/2         \\
with $\theta_{\rm ari}$ & $\mathscr{D}$   & $62.3$ & $62.3$        & $62.5$  & $59.9$          \\
\enddata
\end{deluxetable}
%-----------------------------Table End------------------------------

%-----------------------------Table Start------------------------------
\begin{deluxetable}{ccccccccc}
\tabletypesize{\scriptsize}
%\rotate
\tablecolumns{9}
\tablewidth{0pc}
\tablecaption{Differences between $\theta_{\rm ari}$ and $\theta_{\rm ave}$ and distances $x$ and $d$.\label{tbl:theta_comb}}
\tablehead{No. & Model Name  & $(\theta_{\rm ave}-\theta_{\rm ari})_1$ & $x_1^{\rm a}$ & $d_1^{\rm b}$
& $(\theta_{\rm ave}-\theta_{\rm ari})_2$ & $x_2^{\rm a}$ &  $d_2^{\rm b}$ &  Acpt.  \\
  &  & arcsec &  & kpc & arcsec &  & kpc &
}
\startdata
01 & MRN         &  0.008  & $0.967^{+0.002}_{-0.002}$ & $2.17^{+0.42}_{-0.42}$ &  0.009  & $0.72^{+0.04}_{-0.06}$  & $0.87^{+0.17}_{-0.26}$ &    \\
\noalign{\smallskip}
02 & WD01        &  0.007  & $0.961^{+0.003}_{-0.003}$ & $2.58^{+0.51}_{-0.51}$ &  0.002  & $0.63^{+0.07}_{-0.11}$  & $1.32^{+0.40}_{-0.62}$ &    \\
\noalign{\smallskip}
03 & BARE-GR-S   & -0.009  & $0.968^{+0.002}_{-0.003}$ & $2.10^{+0.40}_{-0.43}$ & -0.027  & $0.71^{+0.04}_{-0.07}$  & $0.91^{+0.18}_{-0.31}$ &    \\
04 & BARE-GR-FG  & -0.004  & $0.968^{+0.003}_{-0.002}$ & $2.10^{+0.40}_{-0.43}$ &  0.008  & $0.71^{+0.04}_{-0.06}$  & $0.91^{+0.18}_{-0.26}$ &    \\
05 & BARE-GR-B   &  0.019  & $0.973^{+0.003}_{-0.002}$ & $1.77^{+0.38}_{-0.35}$ &  0.032  & $0.75^{+0.04}_{-0.05}$  & $0.75^{+0.16}_{-0.20}$ &    \\
\noalign{\smallskip}
06 & BARE-AC-S   &  0.004  & $0.967^{+0.003}_{-0.003}$ & $2.17^{+0.44}_{-0.44}$ & -0.039  & $0.71^{+0.04}_{-0.07}$  & $0.91^{+0.18}_{-0.31}$ &    \\
07 & BARE-AC-FG  & -0.009  & $0.967^{+0.002}_{-0.003}$ & $2.17^{+0.44}_{-0.44}$ &  0.040  & $0.70^{+0.05}_{-0.06}$  & $0.96^{+0.23}_{-0.28}$ &    \\
08 & BARE-AC-B   & -0.019  & $0.972^{+0.002}_{-0.003}$ & $1.83^{+0.36}_{-0.39}$ & -0.038  & $0.75^{+0.03}_{-0.06}$  & $0.75^{+0.12}_{-0.24}$ &    \\
\noalign{\smallskip}
09 & COMP-GR-S   & -0.005  & $0.957^{+0.003}_{-0.004}$ & $2.86^{+0.56}_{-0.59}$ &  0.014  & $0.57^{+0.08}_{-0.12}$  & $1.69^{+0.55}_{-0.83}$ & $\bigcirc$   \\
10 & COMP-GR-FG  & -0.006  & $0.958^{+0.003}_{-0.003}$ & $2.65^{+0.52}_{-0.52}$ &  0.005  & $0.56^{+0.06}_{-0.09}$  & $1.76^{+0.43}_{-0.64}$ & $\bigcirc$   \\
11 & COMP-GR-B   & -0.006  & $0.967^{+0.003}_{-0.003}$ & $2.17^{+0.44}_{-0.44}$ &  0.018  & $0.66^{+0.07}_{-0.09}$  & $1.15^{+0.36}_{-0.46}$ &    \\
\noalign{\smallskip}
12 & COMP-AC-S   &  0.011  & $0.952^{+0.005}_{-0.004}$ & $3.21^{+0.68}_{-0.65}$ & -0.016  & $0.50^{+0.09}_{-0.14}$  & $2.24^{+0.81}_{-1.26}$ & $\surd$  \\
13 & COMP-AC-FG  & -0.003  & $0.945^{+0.004}_{-0.004}$ & $3.70^{+0.73}_{-0.73}$ & -0.013  & $0.53^{+0.07}_{-0.10}$  & $1.99^{+0.56}_{-0.80}$ & $\bigcirc$   \\
14 & COMP-AC-B   &  0.009  & $0.950^{+0.006}_{-0.005}$ & $3.35^{+0.74}_{-0.70}$ & -0.013  & $0.49^{+0.08}_{-0.12}$  & $2.33^{+0.75}_{-1.12}$ & $\surd$   \\
\noalign{\smallskip}
15 & COMP-NC-S   &  0.004  & $0.948^{+0.005}_{-0.005}$ & $3.49^{+0.74}_{-0.70}$ &  0.003  & $0.46^{+0.09}_{-0.13}$  & $2.63^{+0.96}_{-1.38}$ & $\surd$   \\
16 & COMP-NC-FG  & -0.003  & $0.945^{+0.004}_{-0.005}$ & $3.70^{+0.73}_{-0.76}$ & -0.008  & $0.48^{+0.09}_{-0.13}$  & $2.43^{+0.88}_{-1.27}$ & $\surd$   \\
17 & COMP-NC-B   &  0.000  & $0.939^{+0.005}_{-0.005}$ & $4.13^{+0.83}_{-0.83}$ & -0.009  & $0.46^{+0.08}_{-0.12}$  & $2.63^{+0.85}_{-1.27}$ & $\bigcirc$   \\
\noalign{\smallskip}
18 &  XLNW       & -0.012  & $0.968^{+0.002}_{-0.003}$ & $2.10^{+0.41}_{-0.43}$ & -0.030  & $0.70^{+0.04}_{-0.07}$  & $0.96^{+0.18}_{-0.32}$ &    \\
\noalign{\smallskip}
\enddata
\tablecomments{
\textbf{Column Acpt. shows the acceptance of the dust grain models
when compared with the IR distance range $d\in(2.1, 4.2)$ kpc \citep{pel06}.
Those models labeled with $\surd$ are better models,
since $d_i\in (2.1, 4.2)~{\rm kpc},i=1~{\rm and}~2$.
Those models labeled with $\bigcirc$ are also acceptable ones,
since either $d_1$ or $d_2$ is within the distance range,
while the other is consistent with the distance range within $1\sigma$ error,
i.e. $|d_{\rm b}-d_i|\in d_{i, \rm err},i=1~{\rm or}~2$,
where the upper and lower boundary of the distance range
$d_{\rm b}=2.1~{\rm and}~4.2$ kpc, respectively.
The rest of the dust grain models are worse.}}
\tablenotetext{a}{~$x_1$ for 4.5--12.5 arcsec annular halo;
$x_2$ for 27.5--52.5 arcsec annular halo.}
\tablenotetext{b}{~$d_1$ is obatined assuming $\mathscr{D}=63.66\pm11.59$;
$d_2$ is obatined assuming $\mathscr{D}=2.24\pm0.07$ (see Table~\ref{tbl:ari_df}).}
\end{deluxetable}
%-----------------------------Table End------------------------------

%-----------------------------Table Start------------------------------
\begin{deluxetable}{cccc}
\tabletypesize{\footnotesize}
%\rotate
\tablecolumns{4}
\tablewidth{0pc}
\tablecaption{The ratio of cross section of thick dust slab and thin dust slab. \label{tbl:crs_r}}
\tablehead{Model Name  &  $R(\bar x=0.900$)  &  $R(\bar x=0.950$)  &  $R(\bar x=0.990$)}
\startdata
MRN        &  1.00  &  1.00  &  1.09   \\
WD01       &  1.00  &  1.00  &  1.11   \\
XLNW       &  1.00  &  1.00  &  1.02   \\
COMP-GR-S  &  1.00  &  1.00  &  1.07   \\
\enddata
\end{deluxetable}
%-----------------------------Table End------------------------------

\end{document}